\documentclass{elsart}
\usepackage{natbib,epsf}
\newcommand{\beq}{\begin{equation}}
\newcommand{\eeq}[1]{\label{#1} \end{equation}}
\newcommand{\beqar}{\begin{eqnarray}}
\newcommand{\eeqar}[1]{\label{#1} \end{eqnarray}}
\newcommand{\insertplot}[1]{ \begin{center}\leavevmode\epsfysize=7.5cm
\epsfbox{#1}\end{center}} 
\begin{document}


\begin{frontmatter}
\title{Large $p_t$ enhancement from freeze out 
}

\author[Bergen]{V.K. Magas},
\author[Bergen,Frankfurt]{Cs. Anderlik},
\author[Bergen,Frankfurt,Budapest]{L.P. Csernai},
\author[Sao-Paulo]{F. Grassi},
\author[Frankfurt]{W. Greiner},
\author[Sao-Paulo]{Y. Hama},
\author[Rio-de-J]{T. Kodama},
\author[Bergen,Frankfurt]{Zs.I. L\'az\'ar}
and 
\author[Frankfurt]{H. St\"ocker}
\address[Bergen]{Section for Theoretical Physics, Department of Physics\\
University of Bergen, Allegaten 55, 5007 Bergen, Norway}
\address[Frankfurt]{Institut f\"ur Theoretische Physik, Universit\"at Frankfurt\\
Robert-Mayer-Str. 8-10, D-60054 Frankfurt am Main, Germany}
\address[Budapest]{KFKI Research Institute for Particle and Nuclear Physics\\
P.O.Box 49,  1525 Budapest, Hungary}
\address[Sao-Paulo]{Instituto de F\'{\i}sica, Universidade de Sao Paulo\\
CP 66318, 05389-970 S\~ao Paulo-SP, Brazil}
\address[Rio-de-J]{Instituto de F\'{\i}sica, Universidade Federal do Rio de Janeiro\\
CP 68528, 21945-970 Rio de Janeiro-RJ, Brazil}

\begin{abstract}
Freeze out of particles across three dimensional space-time hypersurface 
is discussed in a simple kinetic model. The final momentum distribution of 
emitted particles, for freeze out surfaces with space-like normal, 
shows a non-exponential transverse momentum spectrum. The slope parameter
of the $p_t$ distribution increases with increasing $p_t$, in agreement
with recently measured SPS pion and $h^-$ spectra.
\end{abstract}
\end{frontmatter}

\section{Introduction}
In continuum and
fluid dynamical models, 
particles, which leave the system and reach the detectors, can be
taken into account via freeze out
(FO) or final break-up schemes, where the frozen out particles are formed on
a \mbox{3-dimensional} hypersurface in space-time. Such FO
descriptions are
important ingredients of evaluations of two-particle correlation data, 
\mbox{transverse-,} \mbox{longitudinal-,} radial-, and cylindrical-
flow analyses,
transverse momentum and transverse mass spectra and many other observables.
The FO on a hypersurface is a discontinuity where the pre-FO equilibrated and
interacting matter abruptly changes to non-interacting particles, showing an
ideal gas type of behavior.

The general theory of discontinuities in
relativistic flow was not worked out for a long time, and the 1948 work of A.
Taub \cite{Ta48} discussed discontinuities across propagating hypersurfaces
only (which have a space-like unit normal vector, 
\mbox{$d\hat{\sigma}^\mu d\hat{\sigma}_\mu = -1$).}
Events happening on a propagating, (2 dimensional) surface belong 
to this category.
An overall sudden change in a
finite volume is represented by a hypersurface with a time-like normal,
$d\hat{\sigma}^\mu d\hat{\sigma}_\mu = + 1$.
The freeze out surface is frequently
a surface with time like normal. 
In 1987 Taub's 
approach was generalized
to both types of surfaces \cite{Cs87}, making it possible to take into account
conservation laws exactly across any surface of discontinuity in
relativistic flow.
When the 
EoS is different on the two sides of the freeze out front these
conservation laws yield changing temperature, density, flow velocity across
the front  
\cite{CF74,Bu96,ALC98,AC98,9ath99}.

\section{Conservation laws across idealized freeze out fronts} 
The freeze out 
surface is an idealization of
a layer of finite thickness (of the order of a mean free path or collision 
time) where the frozen-out particles are formed and the interactions
in the matter become negligible. 

To use well-known Cooper-Frye formula \cite{CF74}
\beq
E\frac{dN}{d^3p} = \int f_{FO}(x,p;T,n,u^\nu) \ p^\mu d\sigma_\mu
\eeq{e-cf}
we have to know the post-FO 
distribution
of frozen out particles, $f_{FO}(x,p;T,n,u^\nu)$, which is not known from the 
fluid dynamical model. 
To evaluate measurables we have
to know the correct parameters of the matter {\em after} the FO discontinuity!
The post freeze out distribution need not
be a thermal distribution! In fact $f_{FO}$ should contain only particles
which cross the FO-front outwards, $p^\mu d\hat{\sigma}_\mu >0$, so if $d\hat{\sigma}^\mu$
is space-like this seriously constrains the shape of $f_{FO}$. This
problem was recognized in recent years, and the first suggestions for the
solution were published recently \cite{Bu96,ALC98,AC98,9ath99}.

If we know the pre freeze out baryon current and energy-momentum tensor, 
$N_0^\mu$ and $T_0^{\mu\nu}\ ,$ 
we can calculate locally, across a surface element of normal vector
$d\hat{\sigma}^\mu$
the post freeze out quantities, 
$N^\mu$ and $T^{\mu\nu}$, 
from the relations ~\cite{Ta48,Cs87}:
$
[N^\mu\ d\hat{\sigma}_\mu] = 0 
$
and
$
[T^{\mu\nu}\ d\hat{\sigma}_\mu] = 0 ,
$
where $[A]\equiv A - A_0$.
In numerical calculations the local freeze out surface can be determined most
accurately via self-consistent iteration \cite{Bu96,NL97}.

\section{Freeze out distribution from kinetic theory}
We present a kinetic
model simplified to the limit where we can 
obtain a post FO particle momentum distribution.
Let us assume an infinitely long tube
with its left half ($x<0$) filled with nuclear mater and in the right 
vacuum is maintained. We can remove the dividing wall at $t=0$, and then 
the matter will expand into the vacuum. By continuously removing
particles at the right end of the tube and supplying particles on the 
left end, we can establish a stationary flow in the tube, where the
particles will gradually freeze out in an exponential rarefraction wave
propagating to the left in the matter.  We can move with this front, so
that we describe it from the reference frame of the front (RFF).

We can describe the freeze out kinetics on the r.h.s. of the tube
assuming that we have two components of our momentum 
distribution,  
$f_{free}(x,\vec{p})$  and
$f_{int}(x,\vec{p})$. However, we assume that at $x=0$, $f_{free}$
vanishes exactly and $f_{int}$ is an ideal J\"uttner distribution,
then $f_{int}$ gradually
disappears and $f_{free}$ gradually builds up.

Rescattering within the interacting  component will lead to
re\--ther\-ma\-li\-za\-tion 
and re-equilibration of this component. Thus, the evolution
of the component, $f_{int}$ is determined by drain terms and the
re-equilibration.

We use the  relaxation 
time approximation to simplify the description of the dynamics.
Then 
the two components of the momentum distribution develop according to the
coupled differential equations:
\beq
\begin{array}{rll}
\partial_x f_{int}(x,\vec{p}) dx =& - \Theta(p^\mu d\hat{\sigma}_\mu) 
                                   \frac{\cos \theta_{\vec{p}} }{\lambda}
           f_{int}(x,\vec{p})   dx+
 \\ & & \\
                                  
           &+\left[ f_{eq}(x,\vec{p}) -  f_{int}(x,\vec{p})\right]
           \frac{1}{\lambda'} dx, 
\end{array}
\eeq{kin-2}
\beq
\begin{array}{rll}
\partial_x f_{free}(x,\vec{p}) dx =& + \Theta(p^\mu d\hat{\sigma}_\mu) 
                                   \frac{\cos \theta_{\vec{p}} }{\lambda}
           f_{int}(x,\vec{p})  dx.
\end{array}
\eeq{kin-3}
Here $\cos \theta_{\vec{p}}=p^x/p $ in the RFF frame.
The first (loss) term in eq. (\ref{kin-2}) is an overly simplified
approximation to the model presented in ref. \cite{ALC98}. It expresses the 
fact that particles with momenta orthogonal to the FO surface 
($\cos \theta_{\vec{p}}=1$) leave the system with bigger probability
than particles emitted at an angle.
The interacting component of the momentum distribution, described by 
eq. (\ref{kin-2}), shows the tendency to approach an equilibrated
distribution with a relaxation length  $\lambda' $.  Of course,
due to the energy, momentum and  particle drain, this distribution,
$f_{eq}(x,\vec{p})$ is not the same as the initial J\"uttner distribution,
but its parameters, $n_{eq}(x)$, $T_{eq}(x)$ and $u^\mu_{eq}(x)$,
change as required by the conservation laws.

In this case the change of the conserved quantities 
caused by the particle transfer from component $int$ to component $free$
can be obtained in terms of the distribution functions as:
\beq
d N_i^\mu = 
      -\frac{dx}{\lambda}\int\frac{d^3p}{p_0} p^\mu    
 \Theta(p^\mu d\hat{\sigma}_\mu) \cos \theta_{\vec{p}}
   f_{int}(x,\vec{p})
\eeq{kin-4}
and
\beq
d T_i^{\mu\nu} = 
      -\frac{dx}{\lambda}\int\frac{d^3p}{p_0} p^\mu p^\nu 
 \Theta(p^\mu d\hat{\sigma}_\mu) \cos \theta_{\vec{p}}
f_{int}(x,\vec{p}).
\eeq{kin-5}
Due to the collision or relaxation terms
$T^{\mu\nu}$ and $N^\mu$ change, and this should be considered in the modified 
distribution function $f_{int}(x,\vec{p})$.

\subsection{Immediate re-thermalization limit} 
Let us assume 
that $\lambda' \ll \lambda$, i.e. re-thermalization is much faster
than particles freezing out, or much faster than parameters,
$n_{eq}(x)$, $T_{eq}(x)$ and $u^\mu_{eq}(x)$ change.
Then 
$
f_{int}(x,\vec{p})\approx f_{eq}(x,\vec{p})
$,
for 
$
\lambda'\ll \lambda\,.
$

For $f_{eq}(x,\vec{p})$ we assume the spherical J\"uttner form at
any $x$ including both positive and negative momentum parts 
with parameters $n(x),\ T(x)$ and $u_{RFG}^\mu(x)$.
(Here $u_{RFG}^\mu(x)$ is the actual flow velocity of the
interacting, J\"uttner component, i.e. the velocity of the Rest Frame of the
Gas (RFG) \cite{Bu96}).

In this case the change of conserved quantities due to particle drain or 
transfer can be  evaluated for an infinitesimal $dx$. 
 The changes of the conserved particle currents and
energy-momentum tensor in the RFF, 
eqs. (\ref{kin-4},\ \ref{kin-5}) are given in ref. \cite{ALC98}.
The new parameters of distribution $f_{int}$, 
after moving to the right by $dx$ can be obtained from
$dN_i^\mu$ and $dT_i^{\mu\nu}$.
The differential equation
describing the change of the proper particle density is \cite{ALC98}:
\beq
dn_i(x) =  u_{i,RFG}^\mu(x)\ dN_{i,\mu}(x) \,. 
\eeq{dnx}
Although this covariant
equation is valid in any frame, 
$dN_i^\mu$ are calculated  in the RFF \cite{ALC98}. 

For the re-thermalized interacting component
the change of Eckart's flow velocity is 
given by
\beq
du^\mu_{i,E,RFG}(x)=\Delta^{\mu\nu}_i(x)\  \frac{dN_{i,\nu}(x)}{n_i(x)}\,,
\eeq{duex}
where
$
\Delta^{\mu\nu}_i(x) = g^{\mu\nu} -  u_{i,RFG}^\mu(x)\, u_{i,RFG}^\nu(x)\, 
$\ \ 
is a projector to the plane orthogonal to
$
u^\mu_{i,RFG}(x)  
$, 
while
the change of Landau's flow velocity is
\cite{ALC98}
\beq
du^\mu_{i,L,RFG}(x) = \frac{ 
\Delta^{\mu\nu}_i(x)\ \ dT_{i,\nu\sigma}\ \ u^{\sigma}_{i,RFG}(x)}
{e_i + P_i}.
\eeq{dulx}
Although, 
for the spherical J\"uttner distribution  
the Landau and Eckart flow velocities are the same,
the change of this flow velocity  calculated from the loss of baryon current
and from the loss of energy current are different
$
du^\mu_{i,E,RFG}(x) \ne  du^\mu_{i,L,RFG}(x) \,.
$
This is a clear consequence of the asymmetry caused by the freeze out
process as it was discussed in ref. \cite{ALC98}, i.e., the cut by 
$\Theta(p^\mu d\hat{\sigma}_\mu)$ changes the particle flow 
and energy-momentum flow differently.
This problem does not occur for the freeze out of baryonfree 
plasma, and we have only $du^\mu_{i,L}$. 


\begin{figure}[htb]
	\insertplot{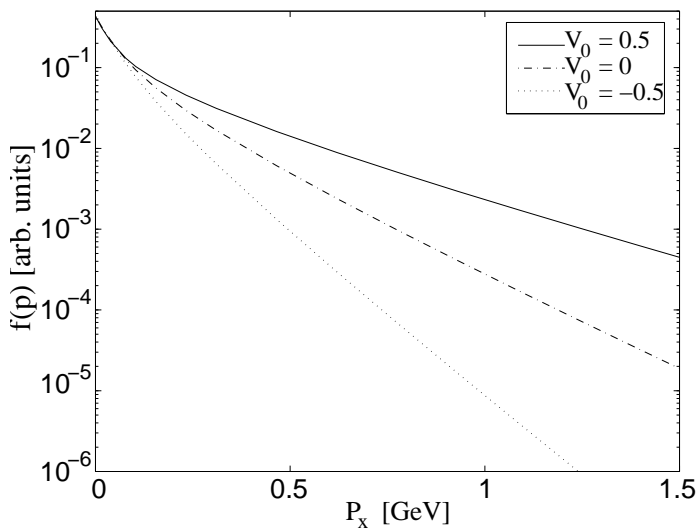}
\caption{The 
local transverse momentum (here $p_x$)
distribution for baryon free, massless gas 
at $p_y = 0$, $x=100 \lambda$ and $T_0 = 130\,$MeV.  
The transverse momentum spectrum is obviously curved
due to the freeze out process, particularly for large
initial flow velocities. The apparent slope parameter
increases with increasing transverse momentum. This behavior agrees
with observed pion transverse mass spectra at SPS \cite{na44,na49}.
}

\label{fig:1}
\end{figure}

The last task is to determine the change of the temperature parameter
of $f_{int}$.  From the relation
$
e\equiv u_\mu T^{\mu\nu} u_\nu
$
we readily obtain the expression for the change of energy density
\beq 
de_i(x) =u_{\mu,i,RFG}(x)\ dT^{\mu\nu}_i(x)\ u_{\nu,i,RFG}(x) \,,
\eeq{dedx}
and from the relation between the energy density and the
temperature (see Chapter 3 in ref.~\cite{Cs94}), we can obtain
the new temperature at $x+dx$.
Fixing these parameters we fully determined the spherical J\"uttner
approximation for $f_{int}$.

The application of this model to the baryonfree and massless gas gives
 the following coupled set of equations:
\beqar
 \frac {d\ln{T}}{dx}& =& -\frac{u_{\mu}\tau^{\mu\nu}u_{\nu}}{4\sigma_{SB}}\,,
 \nonumber \\ & & \\
 \frac {du^{\mu}}{dx}& =& -\frac{3}{4\sigma_{SB}}\left[\tau^{\mu\nu}-u^{\mu}
 u_{\sigma}\tau^{\sigma \nu}\right]u_{\nu}\,. \nonumber
 \eeqar{simmod}
Here we use the EoS, $e=\sigma_{SB} T^4$,\ \ the definition 
\mbox{$dT^{\mu\nu}$ $=$ $-dx\,\tau^{\mu\nu} T^4$,}
\ \ and  $x$ is measured in units of $\lambda$.

Now we can find
the distribution function for the noninteracting, frozen out part of particles 
according to equation (\ref{kin-3}). The results are
shown in Fig. \ref{fig:1}.
We would like to note that now $f_{int}(x,\vec{p})$ does not
tend to the cut J\"uttner distribution in the limit $x \rightarrow \infty$. 
Furthermore, we obtain that $T \rightarrow 0$,
when $x \rightarrow \infty$ \cite{ALC98}. 
So, $f_{int}(x,\vec{p})=\frac{1}{(2\pi\hbar)^3}\exp[(\mu-p^{\nu}u_{\nu})/T] \rightarrow 0$, 
when $x \rightarrow \infty$. Thus, all particles freeze
  out in the present model, but such a physical FO requires 
 infinite distance (or time). This second problem may also be 
 removed by using volume emission model 
 discussed in \cite{9ath99}.

\section{Conclusions} 
In a simple kinetic model we evaluated the freeze out
distribution, $f_{free}(x,p)$, for \linebreak stationary freeze out across 
a surface
with space-like normal vector, $d\hat{\sigma}^\mu d\hat{\sigma}_\mu < 0$.
In this model particles penetrating
the surface outwards were allowed to freeze out with a probability
$\sim \cos \theta_{\vec{p}}$, and the remaining  
 interacting component is assumed to be instantly
re-thermalized.  The three parameters of the interacting component,
$f_{int}$, are obtained in each time step.  The density of the interacting
component gradually decreases and disappears, the flow velocity also
decreases and the energy density decreases.  The temperature, as a
consequence of the gradual change in the emission mechanism, 
gradually decreases.

The arising post freeze out distribution, $f_{free}$ is a 
superposition of cut J\"uttner type of components, from a series of gradually
slowing down J\"uttner distributions. This leads to a 
final momentum distribution, with a more dominant peak at zero
momentum
and a forward halo, Fig. \ref{fig:1}.  
In this rough model a large fraction ($\sim 95\%$)
of the matter is frozen out by $x=3 \lambda$, thus, the distribution
$f_{free}$ at this distance can be considered as a first estimation
of the post freeze out distribution.  One should also keep in mind that 
the model presented here  does not have realistic behavior in the limit 
$x \longrightarrow \infty$, due to its one dimensional character.

These studies indicate that more attention should be paid
to the final freeze out process, because a realistic freeze
out description may lead to large $p_t$ enhancement \cite{na44,na49}
as the considerations above indicate (Fig. \ref{fig:1}). 
For accurate estimates
more realistic models should be used.
In case of rapid hadronization of QGP and 
simultaneous freeze out, the idealization of a freeze out hypersurface
may be justified, however, an accurate determination of the post freeze
out hadron momentum distribution would require a nontrivial dynamical
calculation.

\ack
This work is supported in part by the Research Council of Norway, PRONEX
(contract no. 41.96.0886.00), FAPESP (contract no. 98/2249-4) and CNPq.
Cs. Anderlik, L.P. Csernai and Zs.I. L\'{a}z\'{a}r are thankful for the
hospitality extended to them by the Institute for Theoretical Physics of the
University of Frankfurt where part of this work was done. L.P. Csernai is
grateful for the Research Prize received from the Alexander von Humboldt
Foundation.

%

\vfill\eject
\end{document}